\documentclass[12pt]{article}
\usepackage[dvips]{graphicx}
\usepackage{epsfig,amsmath}

\begin{document}
\thispagestyle{empty}
\begin{center} 
{\Large\bf Statistical parton distributions,\\

 TMD, positivity and all that\footnote{Invited talk at the XIV Advanced Research Workshop on High Energy Spin Physics, "`DSPIN-2011"', JINR, Dubna, Russia, September 20 - 24, 2011, to appear in the Proceedings}}\\

\vskip 1.4cm
{\bf Jacques Soffer}
\vskip 0.3cm
{\it Physics Department, Temple University},\\ 
{\it Barton Hall, 1900 N, 13th Street},\\
{\it Philadelphia, PA 19122-6082, USA}\\
\end{center}
\vskip 1.5cm
\begin{center}
{\bf Abstract}
\end{center}

We briefly recall the main physical features of the parton distributions in the quantum statistical picture of the nucleon. Some predictions from a next-to-leading order QCD analysis are successfully compared to recent unpolarized and polarized experimental results. We will discuss the extension to the transverse momentum dependence of the parton distributions and its relevance for semiinclusive deep inelastic scattering. Finally, we will present some new positivity constraints for spin observables and their implications for parton distributions.\\


\vskip 1.4cm
\newpage

\section{A short review on the statistical approach}
Let us first recall some of the basic ingredients for building up the parton distribution functions (PDF) in the statistical approach, as oppose to the standard polynomial type
parametrizations, based on Regge theory at low $x$ and counting rules at large $x$.
The fermion distributions are expressed by the sum of two terms \cite{bbs1}, the first one, 
a quasi Fermi-Dirac function, for a given helicity and flavor, and the second one, a flavor and helicity independent diffractive
contribution equal for light quarks. So we have, at the input energy scale $Q_0^2=4 \mbox{GeV}^2$,
\begin{equation}
xq^h(x,Q^2_0)=
\frac{AX^h_{0q}x^b}{\exp [(x-X^h_{0q})/\bar{x}]+1}+
\frac{\tilde{A}x^{\tilde{b}}}{\exp(x/\bar{x})+1}~,
\label{eq1}
\end{equation}
\begin{equation}
x\bar{q}^h(x,Q^2_0)=
\frac{{\bar A}(X^{-h}_{0q})^{-1}x^{2b}}{\exp [(x+X^{-h}_{0q})/\bar{x}]+1}+
\frac{\tilde{A}x^{\tilde{b}}}{\exp(x/\bar{x})+1}~.
\label{eq2}
\end{equation}
Notice the change of sign of the potentials
and helicity for the antiquarks.
The parameter $\bar{x}$ plays the role of a {\it universal temperature}
and $X^{\pm}_{0q}$ are the two {\it thermodynamical potentials} of the quark
$q$, with helicity $h=\pm$. It is important to remark that the diffractive contribution 
occurs only in the unpolarized distributions $q(x)= q_{+}(x)+q_{-}(x)$ and it is absent in the valence $q_v(x)= q(x) - \bar {q}(x)$ and in the helicity
distributions $\Delta q(x) = q_{+}(x)-q_{-}(x)$ (similarly for antiquarks).
The {\it eight} free parameters\footnote{$A=1.74938$ and $\bar{A}~=1.90801$ are
fixed by the following normalization conditions $u-\bar{u}=2$, $d-\bar{d}=1$.}
in Eqs.~(\ref{eq1},\ref{eq2}) were
determined at the input scale from the comparison with a selected set of
very precise unpolarized and polarized Deep Inelastic Scattering (DIS) data \cite{bbs1}. They have the
following values
\begin{equation}
\bar{x}=0.09907,~ b=0.40962,~\tilde{b}=-0.25347,~\tilde{A}=0.08318,
\label{eq3}
\end{equation}
\begin{equation}
X^+_{0u}=0.46128,~X^-_{0u}=0.29766,~X^-_{0d}=0.30174,~X^+_{0d}=0.22775~.
\label{eq4}
\end{equation}
For the gluons we consider the black-body inspired expression
\begin{equation}
xG(x,Q^2_0)=
\frac{A_Gx^{b_G}}{\exp(x/\bar{x})-1}~,
\label{eq5}
\end{equation}
a quasi Bose-Einstein function, with $b_G=0.90$, the only free parameter
\footnote{In Ref.~\cite{bbs1} we were assuming that, for very small $x$,
$xG(x,Q^2_0)$ has the same behavior as $x\bar q(x,Q^2_0)$, so we took $b_G = 1
+ \tilde b$. However this choice leads to a too much rapid rise of the gluon
distribution, compared to its recent  determination from HERA data, which
requires $b_G=0.90$.}, since $A_G=20.53$ is determined by the momentum sum
rule.
 We also assume that, at the input energy scale, the polarized gluon 
distribution vanishes, so $x\Delta G(x,Q^2_0)=0$. For the strange quark distributions, the simple choice made in Ref. \cite{bbs1}
was greatly improved in Ref. \cite{bbs2}. More recently, new tests against experimental (unpolarized and
polarized) data turned out to be very satisfactory, in particular in hadronic
collisions, as reported in Refs.~\cite{bbs3,bbs4}.\\
An interesting point concerns the behavior of the ratio $d(x)/u(x)$, 
which depends on the mathematical properties of the ratio of two Fermi-Dirac 
factors, outside the region dominated by the diffractive contribution. 
So for $x>0.1$, this ratio is expected to decrease faster for 
$X_{0d}^+ - \bar x < x < X_{0u}^+ + \bar x$ and then above, for 
$x > 0.6$, it flattens out.\\
This change of slope is clearly visible in Fig.~1 ({\it Left}), with a very 
little $Q^2$ dependence. Note that our prediction for the large $x$ behavior,
differs from most of the current literature, namely $d(x)/u(x) \to 0$
for $x \to 1$, but we find $d(x)/u(x) \to 0.16$ near the value $1/5$,
a prediction
originally formulated in Ref.~\cite{FJ}.
This is a very challenging question, since the very high-$x$ region remains
poorly known.
To continue our tests of the unpolarized parton distributions, we must come 
back to the important question of the flavor asymmetry of the light
antiquarks. Our determination of $\bar u(x,Q^2)$ and
$\bar d(x,Q^2)$ is perfectly consistent with the violation of the Gottfried
sum rule, for which we found the value $I_G= 0.2493$ for $Q^2=4\mbox{GeV}^2$.
Nevertheless there remains an open problem with the $x$ distribution
of the ratio $\bar d(x)/\bar {u}(x)$ for $x \geq 0.2$.
According to the Pauli principle, this ratio is expected to remain above 1 for any value of
$x$. However, the E866/NuSea Collaboration \cite{E866} has
released the final results corresponding to the analysis of their full
data set of Drell-Yan yields from an 800 GeV/c proton beam on hydrogen
and deuterium targets and they obtain the ratio, for $Q^2=54\mbox{GeV}^2$, 
$\bar d(x)/\bar {u}(x)$ shown in Fig.~1 ({\it Right}). 
Although the errors are rather large in the high-$x$ region,
the statistical approach disagrees with the trend of the data.
Clearly by increasing the number of free parameters, it
is possible to build up a scenario which leads to the drop off of
this ratio for $x\geq 0.2$.
For example this was achieved in Ref. \cite{Sassot}, as shown 
by the dashed curve in Fig.~1 ({\it Right}). There is no such freedom in the statistical
approach, since quark and antiquark distributions are strongly related. On the experimental side, there are now new
opportunities for extending the $\bar d(x)/\bar {u}(x)$ measurement to larger $x$ up to $x=0.7$, 
with the upcoming E906 experiment at the 120 GeV Main Injector at Fermilab \cite{E906} and a proposed
experiment at the new 30-50 GeV proton accelerator at J-PARC \cite{JPARC}.\\
Analogous considerations can be made for the corresponding helicity 
distributions, whose most recent determinations are shown in Fig.~2 ({\it Left}).
By using a similar argument as above, the ratio $\Delta u(x)/u(x)$ 
is predicted to have a rather fast increase in the $x$ range 
$(X^-_{0u}-\bar{x},X^+_{0u}+\bar{x})$
and a smoother behaviour above, while $\Delta d(x)/d(x)$, which is negative,
has a fast decrease in the $x$ range $(X^+_{0d}-\bar{x},X^-_{0d}+\bar{x})$ 
and a smooth one above. This is exactly the trends displayed in 
Fig.~2 ({\it Right}) and our predictions are in perfect agreement
with the accurate high-$x$ data. We note the behavior near $x=1$, another typical property of the statistical
approach, is also at variance with predictions of the current literature. 
The fact that $\Delta u(x)$ is more concentrated in the higher $x$ region than
$\Delta d(x)$, accounts for the change of sign of $g^n_1(x)$, which becomes
positive for $x>0.5$, as first observed at Jefferson Lab \cite{JLab04}.\\
Concerning the light antiquark helicity distributions, the statistical 
approach imposes a strong relationship to the corresponding quark helicity
distributions. In particular, it predicts $\Delta \bar u(x)>0$ and $\Delta \bar
d(x)<0$, with almost the same magnitude, in contrast with the
simplifying assumption $\Delta \bar u(x)=\Delta \bar d(x)$, often adopted in
the literature. According to the COMPASS experiment 
at CERN \cite{compass}, $\Delta \bar u(x) + \Delta \bar d(x) \simeq 0$, in agreement with our prediction.
\section{The TMD extension}
We now turn to another important aspect of the statistical PDF and very briefly discuss
a new version of the extension to the transverse momentum dependence (TMD).
In Eqs.~(\ref{eq1},\ref{eq2}) the multiplicative factors $X^{h}_{0q}$ and
$(X^{-h}_{0q})^{-1}$ in
the numerators of the non-diffractive parts of $q$'s and $\bar{q}$'s
distributions, imply a modification
of the quantum statistical form, we were led to propose in order to agree with
experimental data. The presence of these multiplicative factors was justified
in our earlier attempt to generate the TMD \cite{bbs5}, but it was not properly done and 
a considerable improvement was achieved recently \cite{bbs6}. We have introduced some thermodynamical
potentials $Y^h_{0q}$, associated to the quark transverse momentum $k_T$, and
related to $X^{h}_{0q}$ by the simple relation
$\mbox{ln}(1+\exp[Y^h_{0q}])=kX^h_{0q}$.
We were led to choose $k=3.05$ and this method involves another parameter
$\mu^2$, which plays the role of the temperature for the transverse degrees of
freedom and whose value was determined by the transverse energy sum rule.
 We have calculated the $p_T$ dependence of semiinclusive DIS cross sections and double
longitudinal spin asymmetries, taking into account the effects of the
Melosh-Wigner rotation, for $\pi^{\pm}$ production by using this
 set of TMD statistical parton distributions and another set coming from the relativistic
covariant approach \cite{zav}. Both sets do not satisfy the usual factorization
assumption of the dependence in $x$ and $k_T$ and they lead to different results, which can be compared
to recent experimental data from CLAS at JLab, as shown on Figs.~3.
\section{Positivity bounds}
Spin observables for any particle reaction, contain some unique information which allow a deeper understanding of the
nature of the underlying dynamics and this is very usefull to check the validity of theoretical assumptions. We 
emphasize the relevance of positivity in spin physics, which puts non-trivial model independent constraints on spin
observables. If one, two or several observables are measured, the constraints can help to decide which new observable will provide
the best improvement of knowledge. Different methods can be used to establish these constraints and they have been presented together with many interesting cases in a review article \cite{aerst}.
For lack of space, here we will only briefly discuss some new results obtained very recently \cite{ks1, ks2}.\\
Let us consider the inclusive reaction of the type  $A(\mbox{spin 1/2})+B(\mbox{spin 1/2})\to C+X$, 
where both initial spin $1/2$ particles can be in any possible directions and no polarization is observed in the final state. The spin-dependent corresponding cross section $\sigma\left(P_a, P_b\right)={\rm Tr}\left(M\rho\right)$, can be defined through the $4\times 4$ cross section matrix $M$ and the spin density matrix $\rho$, 
where $P_a$, $P_b$ are the spin unit vectors of $A$ and $B$, $\rho=\rho_a\otimes \rho_b$ is the spin density matrix with $\rho_a=(I_2+P_a\cdot \vec{\sigma}_a)/2$, and similar for $\rho_b$. Here $I_2$ is the $2\times 2$ unit matrix, and $\sigma=(\sigma_x, \sigma_y, \sigma_z)$ stands for the $2\times 2$ Pauli matrices.
$M$ can be parametrized in terms of 8 parity-conserving asymmetries and 8 parity-violating asymmetries. The crucial point is that $M$ is a Hermitian and positive matrix and this allows to 
derive some positivity conditions.
Since one of the necessary conditions for a Hermitian matrix to be positive definite is that all the diagonal matrix elements has to be positive $M_{ii}\geq 0$, we thus derive
$1\pm A_{NN}\geq \left|A_{aN}\pm A_{bN}\right|$, valid in full generality, for both parity-conserving and parity-violating processes, where $A_N$ denotes the single transverse spin asymmetry and $A_{NN}$ the double transverse spin asymmetry.
In the case $p^\uparrow+p^\uparrow\to C+X$ where the initial particles are identical, we have $A_{aN}(y)=-A_{bN}(-y)$. Using this relation , one obtains,
$1 \pm A_{NN} (y) \geq \left|A_N (y)\mp A_N(-y)\right|$. This is an interesting result which, can be used, in principle, with available data on $A_N$ for $\pi^{\pm}$, $K^{\pm}$, $\pi^0$, $\eta$ production, to put
 some non trivial contraints on $A_{NN}(y)$.\\
Let us now study the implications of the above relation for the parity-violating process $p^\uparrow+p^\uparrow\to W^{\pm}+X$. Since
$A_{NN}\approx 0$, to a very good approximation, it reduces to
$1/2 \geq |A_N (y = 0)|$, 
to be compared with the usual trivial bound $1 \geq |A_N (y = 0)|$.

The TMD quark distribution in a transversely polarized hadron can be expanded as
$f_{q/h^\uparrow}(x,\mathbf{k}_{\perp},\vec{S})
\equiv 
f_{q/h}(x,k_{\perp}) +  
\frac{1}{2}\Delta^N f_{q/h^\uparrow}(x,k_\perp)\,
\vec{S}\cdot \left(\hat{p}\times \hat{\mathbf{k}}_\perp \right)$, 
where $\hat{p}$ and $\hat{\mathbf{k}}_\perp$ are the unit vectors of $\vec{p}$ and 
$\mathbf{k}_\perp$, respectively. $f_{q/h}(x,k_\perp)$ is the spin-averaged TMD distribution, and $\Delta^N f_{q/h^\uparrow}(x,k_\perp)$ is the Sivers function. There is a trivial positivity bound for the Sivers functions which reads $|\Delta^N f_{q/h^\uparrow}(x,k_\perp)|\leq 2f_{q/h}(x,k_{\perp})$. Since $A_N$ is directly expressed in terms of $\Delta^N f_{q/h^\uparrow}(x,k_\perp)$, this trivial bound can be improved as shown in Ref. \cite{ks2}.\\
In the helicity basis it is easy to obtain the explicit form of $M$ and now from $M_{ii}\geq 0$, we have
$1\pm A_{LL}(y)\geq|A_{aL}(y)\pm A_{bL}(y)|$, where $A_L$ denotes the single helicity asymmetry and $A_{LL}$ the double double asymmetry. It is important to note that for identical initial particles scattering, one has
$A_{aL}(y)=A_{bL}(-y)$, so one gets $1\pm A_{LL}(y)\geq|A_{L}(y)\pm A_{L}(-y)|$.
These bounds should be tested in RHIC experiments for $W^{\pm}$ or $Z^0$ production in longitudinal $pp$ collisions, $\vec{p}\vec{p}\to W^{\pm}/Z^0+X$. In perturbative QCD formalism, at leading-order and restricting to only up and down quarks, one has simple expressions for the single and double helicity asymmetries, involving only quark helicity distributions. The statistical
PDF satisfy the positivity bound. Finally at $y=0$, since $A_{LL}(0)$ is expected to be very small, the bound implies $A_{L}(0)\leq
1/2$, a remarquable simple result which must be satisfied by future experimental data.\\

{\bf Acknowledgments}\\
I am grateful to the organizers of DSPIN2011 for their warm hospitality at JINR and for their invitation to present this talk. My special thanks go to Prof. A.V. Efremov for providing a full financial support and for making, once more, this meeting so successful.

\newpage
\begin{figure}[htb]
  \begin{minipage}{7.0cm}
    \vspace*{+10mm}
    \hspace*{-20mm}
  \epsfig{figure=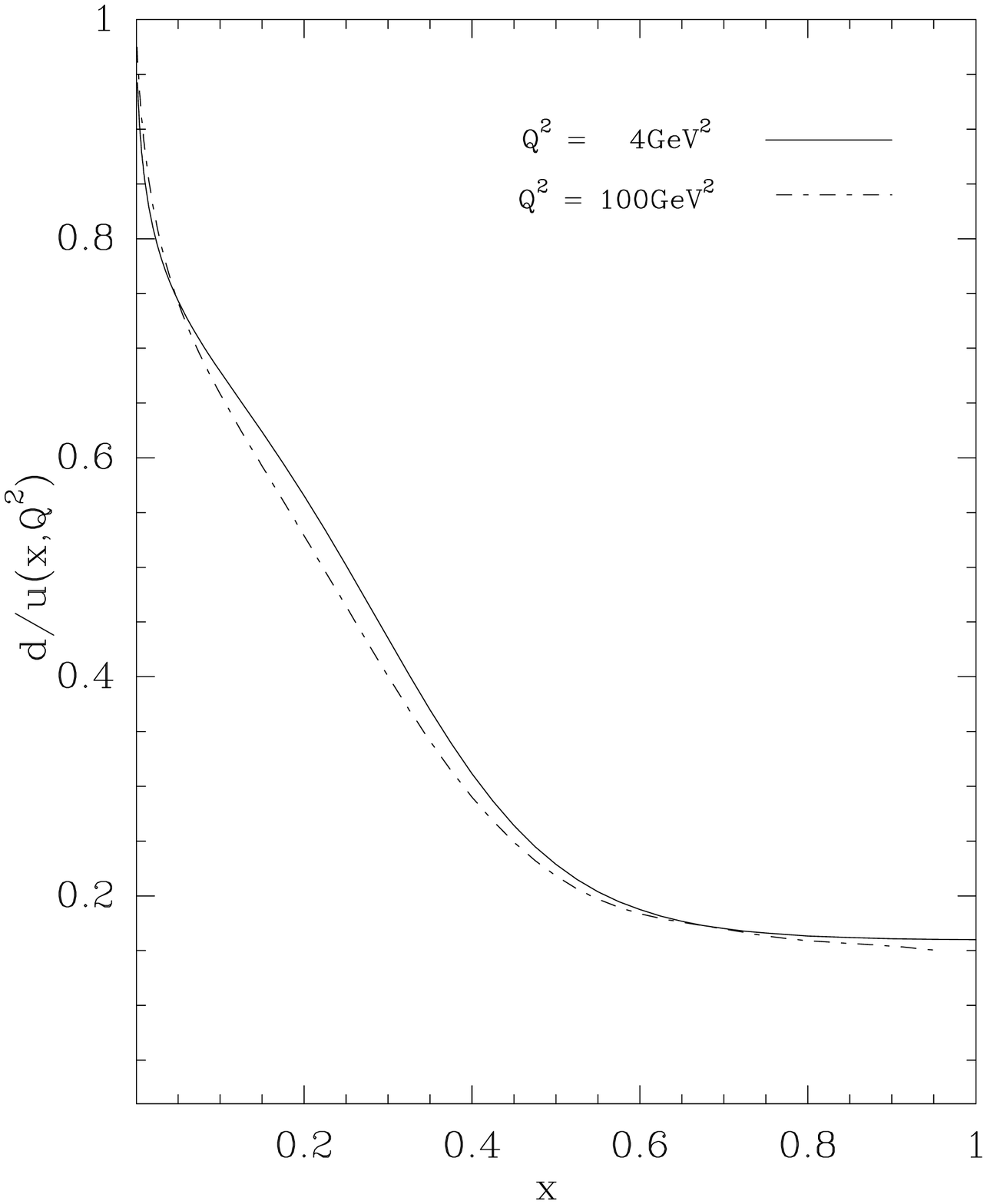,width=8.5cm}
  \end{minipage}
  \vspace*{+3mm}
    \begin{minipage}{7.0cm}
    \hspace*{-10mm}
  \epsfig{figure=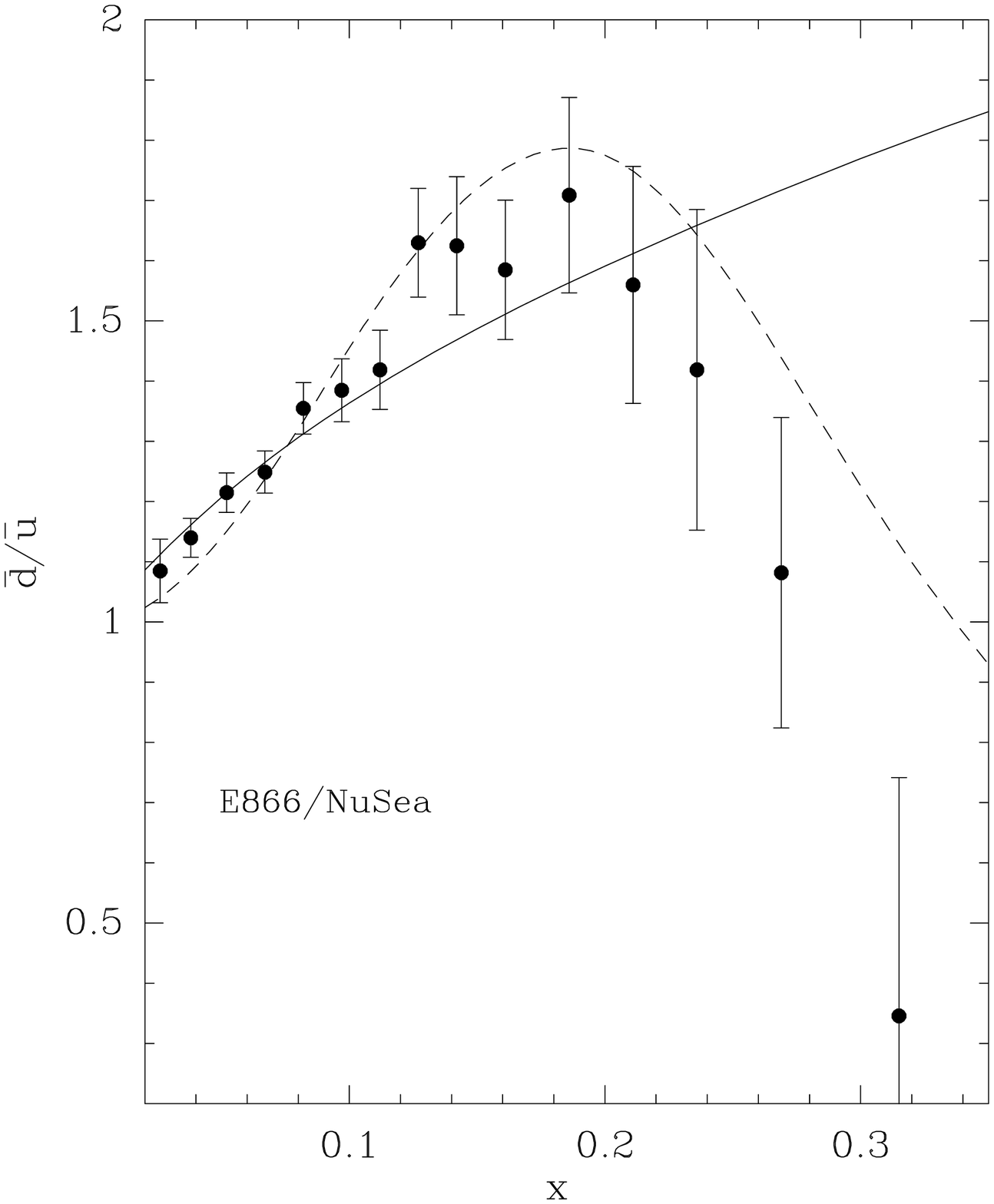,width=8.0cm}
  \vspace*{-10mm}
  \end{minipage}\\
  \caption{{\it Left} : The ratio $d(x)/u(x)$ as function of $x$ for $Q^2 = 4\mbox{GeV}^2$ 
(solid line) and $Q^2 =100\mbox{GeV}^2$ (dashed-dotted line). {\it Right} : Comparison of the data on $\bar d / \bar u (x,Q^2)$ from E866/NuSea
at $Q^2=54\mbox{GeV}^2$
\cite{E866}, with the prediction of the statistical model (solid curve) 
and the set 1 of the parametrization proposed in Ref. \cite{Sassot}
(dashed curve).}
\label{fi:doveru}
\end{figure}
\newpage

\begin{figure}[htb]
  \begin{minipage}{7.0cm}
    \vspace*{+10mm}
    \hspace*{-10mm}
  \epsfig{figure=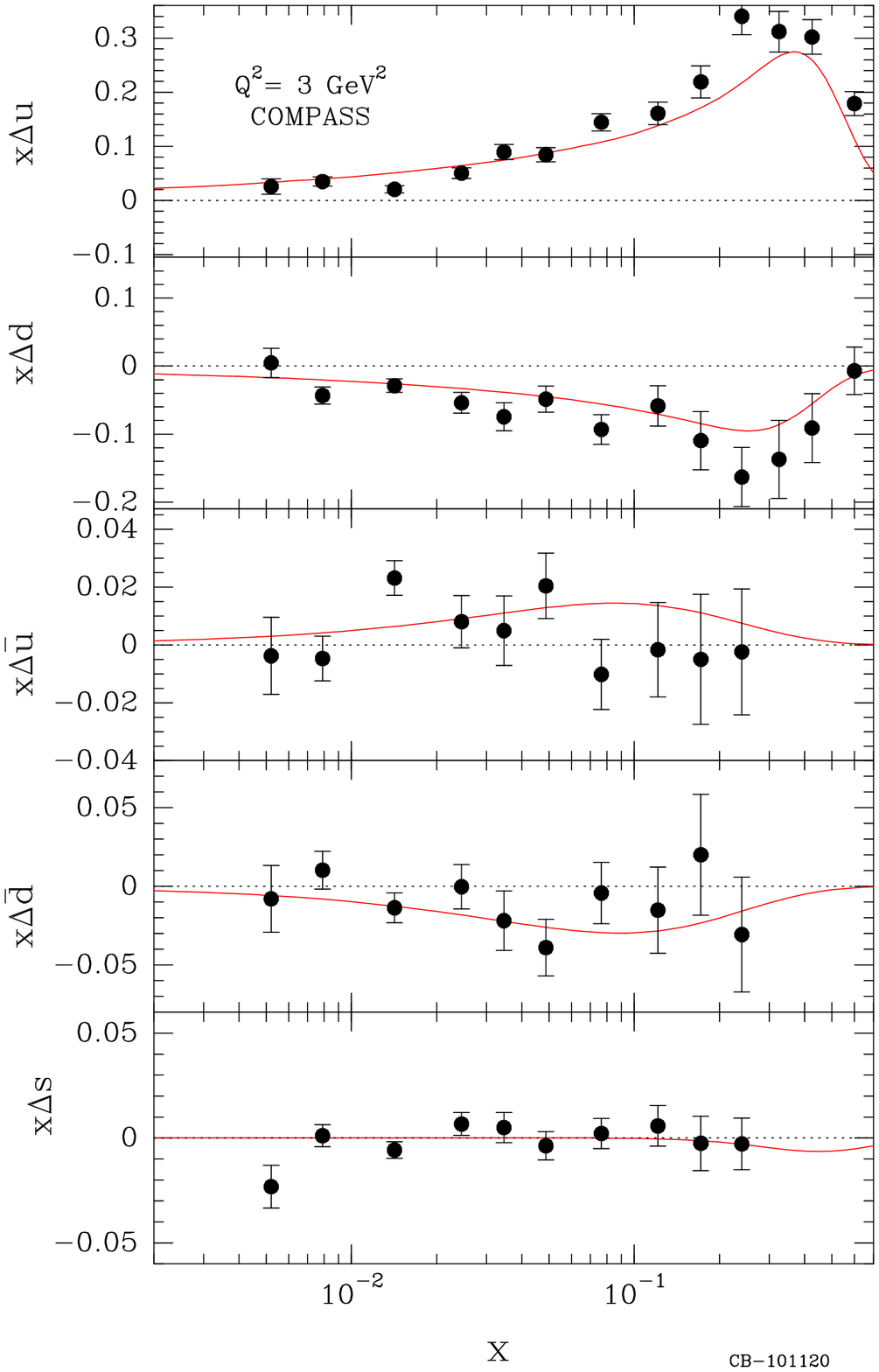,width=8.5cm}
  \end{minipage}
  \vspace*{-3mm}
    \begin{minipage}{7.0cm}
    \hspace*{-10mm}
  \epsfig{figure=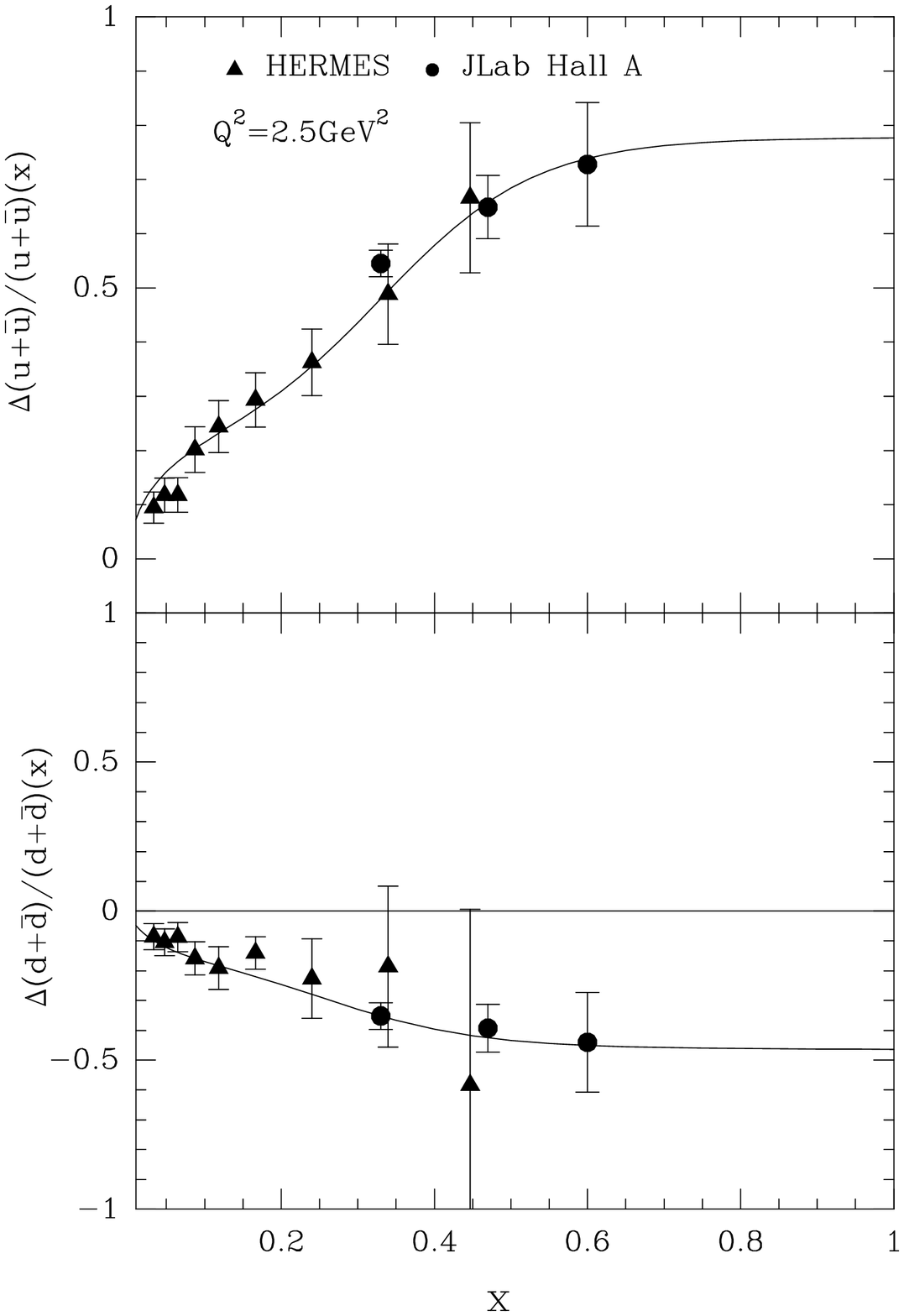,width=9.3cm}
  \vspace*{-25mm}
  \end{minipage}\\
  \caption{{\it Left} : Quark and antiquark helicity distributions as a function of $x$ for $Q^2 = 3\mbox{GeV}^2$.
 Data from COMPASS \cite{COMPASS}. The curves are predictions from the statistical approach. {\it Right} :  Ratios $(\Delta u + \Delta \bar u)/(u + \bar u)$ and 
$(\Delta d + \Delta \bar d)/(d + \bar d)$ as a function of $x$.
Data from Hermes for $Q^2 = 2.5\mbox{GeV}^2$ \cite{herm99} and
a JLab Hall A experiment \cite{JLab04}. The curves are predictions from the statistical approach}
\label{f2:polar}
\end{figure}
\newpage
\begin{figure}[htb]
  \begin{minipage}{7.0cm}
    \vspace*{+10mm}
    \hspace*{-10mm}
  \epsfig{figure=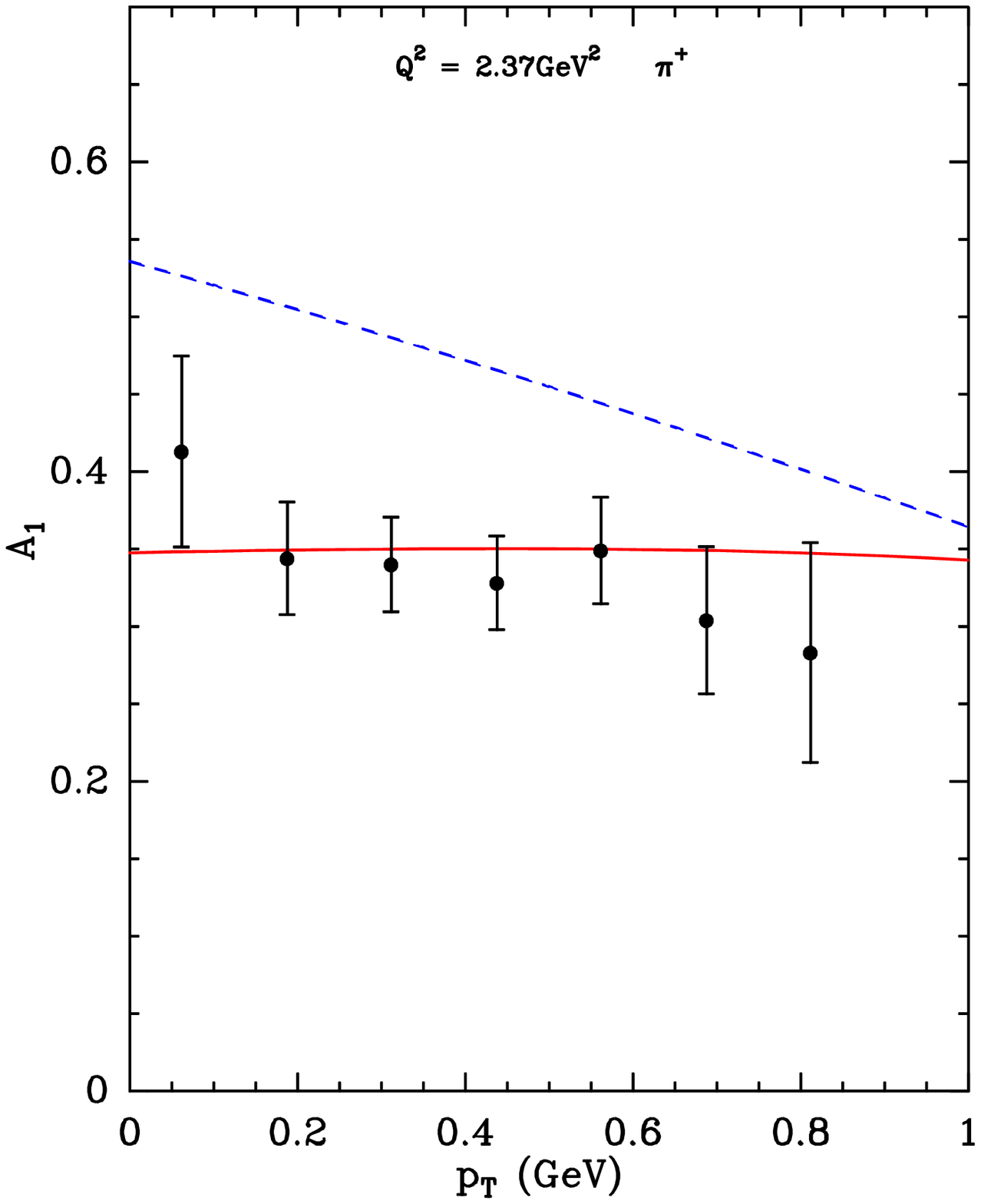,width=9.0cm}
  \end{minipage}
  \vspace*{-3mm}
    \begin{minipage}{7.0cm}
     \vspace*{-10mm}
    \hspace*{-10mm}
  \epsfig{figure=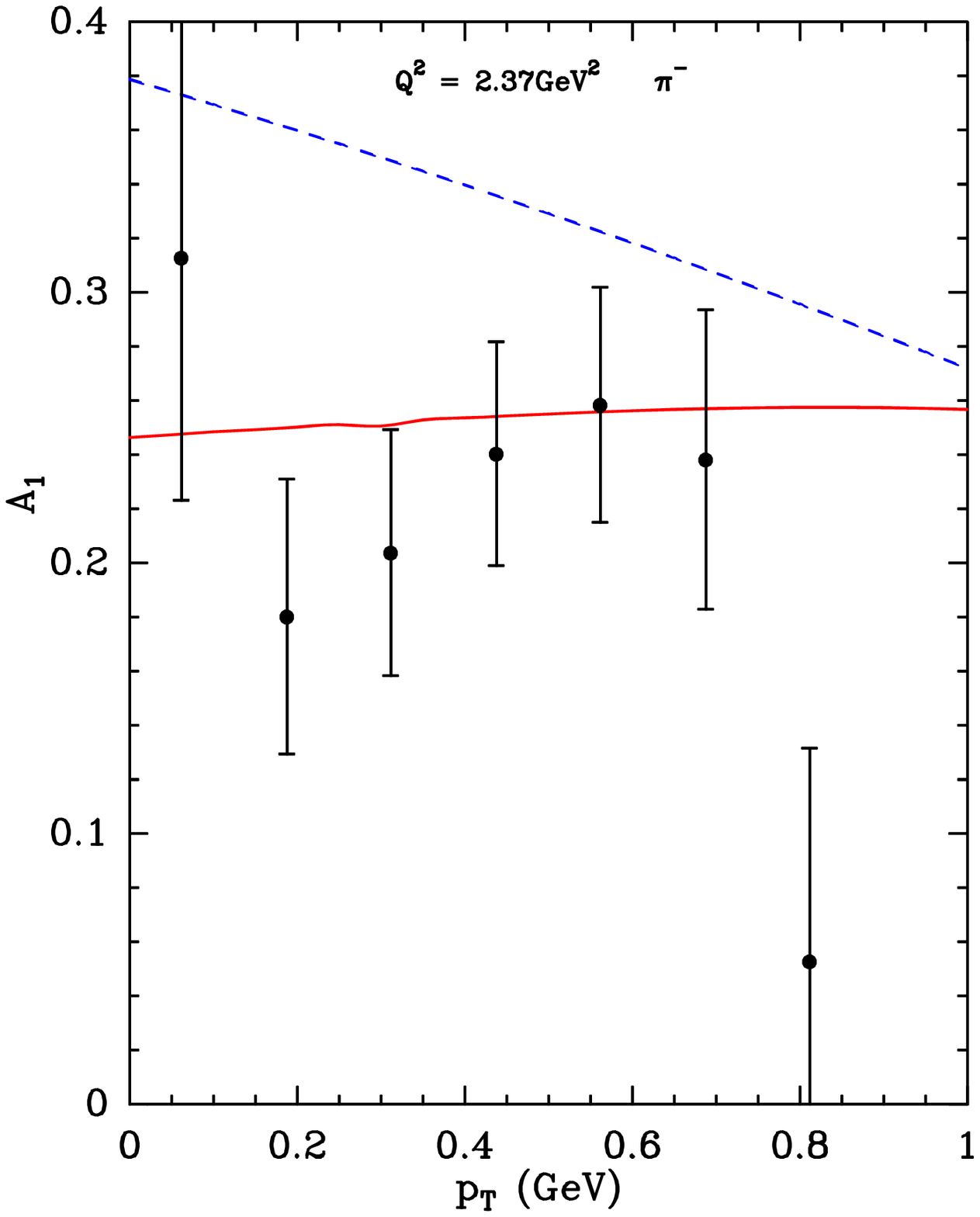,width=9.0cm}
  \vspace*{-25mm}
  \end{minipage}\\
  \caption{{\it Left} : The double longitudinal-spin asymmetry $A_1$ for $\pi^+$ production on a proton target, versus the $\pi^+$ momentum $p_T$, compared to the JLab data Ref.~\cite{clas2}. The solid lines are the
results from the TMD statistical distributions \cite{bbs6} and the dashed lines correspond to
the relativistic covariant distributions \cite{zav} {\it Right} : Same for $\pi^-$.}
\label{f2:polar}
\end{figure}

\end{document}